\documentclass[12pt]{article}  
\usepackage{amssymb,mathrsfs,epsfig,color, url}
\usepackage{amsmath}
\usepackage{graphicx}
\usepackage{slashed}
\usepackage{color}
\usepackage[colorlinks=true
,urlcolor=blue
,citecolor=blue
,linkcolor=blue
,pagecolor=blue
,linktocpage=true
,pdfproducer=medialab
]{hyperref}
\usepackage{epsfig}
\usepackage{subfigure}
\usepackage{cite}

\def\beq{\begin{equation}}
\def\eeq{\end{equation}}
\def\to{\rightarrow}

\def\bsg{\ifmmode B\to X_s\gamma\else $B\to X_s\gamma$\fi}
\def\bsll{\ifmmode B\to X_s\ell^+\ell^-\else $B\to X_s\ell^+\ell^-$\fi}
\def\bstt{\ifmmode B\to X_s\tau^+\tau^-\else $B\to X_s\tau^+\tau^-$\fi}
\def\shat{\ifmmode \hat{s}\else $\hat{s}$\fi}

\def\EmissT{\not \! \!  E_{T}}

\newcommand{\newc}{\newcommand}

\newc{\lcal}{\int {\cal L}dt}
 
\newc{\LSP}{{\chi^0_1}}
\newc{\stauR}{{\tilde \tau_R}}
\newc{\stau}{{\tilde \tau_1}}
\newc{\mstop}{m_{\tilde{t}}}
\newc{\mHpm}{m_{H^\pm}}
\newc{\gsim}{\lower.7ex\hbox{$\;\stackrel{\textstyle>}{\sim}\;$}}
\newc{\lsim}{\lower.7ex\hbox{$\;\stackrel{\textstyle<}{\sim}\;$}}
\newc{\ie}{{\it i.e.}}          
\newc{\etal}{{\it et al.}}
\newc{\eg}{{\it e.g.}}          
\newc{\kev}{\hbox{\rm\,keV}}            
\newc{\mev}{\hbox{\rm\,MeV}}            
\newc{\gev}{\hbox{\rm\,GeV}}            
\newc{\tev}{\hbox{\rm\,TeV}}
\newc{\xpb}{\hbox{\rm\, pb}}
\newc{\xfb}{\hbox{\rm\, fb}}

\newc{\mtop}{m_t}
\newc{\mbot}{m_b}
\newc{\mz}{M_Z}
\newc{\mw}{M_W}
\newc{\alphasmz}{\alpha_s(M_Z)}
\newc{\swsq}{\sin^2\theta_W}
\newc{\cwsq}{\cos^2\theta_W}
\newc{\tw}{\tan\theta_W}
\newc{\cw}{\cos\theta_W}
\newc{\sw}{\sin\theta_W}
\newc{\BR}{\hbox{\rm BR}}
\newc{\zbb}{Z\to b\bar}
\newc{\Gb}{\Gamma (Z\to b\bar b)}
\newc{\Gh}{\Gamma (Z\to \hbox{\rm hadrons})}
\newc{\rbsm}{R_b^\hbox{\rm sm}}
\newc{\rbsusy}{R_b^\hbox{\rm susy}}
\newc{\drb}{\delta R_b}

\newc{\sgn}{\mbox{sgn}}

\def\beqa{\begin{eqnarray}}
\def\eeqa{\end{eqnarray}}
\def\slash#1{\not  \! \! {#1}}

\def\beq{\begin{equation}}
\def\eeq{\end{equation}}
\def\bea{\begin{eqnarray}}
\def\eea{\end{eqnarray}}
\def\slashchar#1{\setbox0=\hbox{$#1$}           % set a box for #1
   \dimen0=\wd0                                 % and get its size 
   \setbox1=\hbox{/} \dimen1=\wd1               % get size of /
   \ifdim\dimen0>\dimen1                        % #1 is bigger 
      \rlap{\hbox to \dimen0{\hfil/\hfil}}      % so center / in box 
      #1                                        % and print #1
   \else                                        % / is bigger 
      \rlap{\hbox to \dimen1{\hfil$#1$\hfil}}   % so center #1
      /                                         % and print /
   \fi}                                         %
\catcode`@=11
\long\def\@caption#1[#2]#3{\par\addcontentsline{\csname
  ext@#1\endcsname}{#1}{\protect\numberline{\csname
  the#1\endcsname}{\ignorespaces #2}}\begingroup
    \small
    \@parboxrestore
    \@makecaption{\csname fnum@#1\endcsname}{\ignorespaces #3}\par
  \endgroup}
\catcode`@=12

\begin{document}

\begin{center}
\vspace{1cm}

{
\bf\LARGE
%Prediction for SUSY Spectrum  \\[4mm]
%The Upper Bounds on Sfermion Masses  \\[4mm]
The PeV-Scale Split Supersymmetry from 
Higgs Mass and Electroweak Vacuum Stability  }
\\[10mm]
{\large Waqas~Ahmed$^{\,\star}$} \footnote{E-mail: \texttt{waqasmit@itp.ac.cn}},
{\large Adeel Mansha$^{\,\star}$} \footnote{E-mail: \texttt{adeelmansha@itp.ac.cn}},
{\large Tianjun~Li$^{\,\star\,\heartsuit}$} \footnote{E-mail: \texttt{tli@itp.ac.cn}},\\[1mm]
{\large Shabbar~Raza$^{\,\ast}$} \footnote{E-mail: \texttt{shabbar.raza@fuuast.edu.pk}},
{\large Joydeep Roy$^{\,\star}$}
\footnote{E-mail: \texttt{jdroy@itp.ac.cn}},
{\large Fang-Zhou Xu $^{\,\star\,\diamond}$} \footnote{E-mail: \texttt{xfz14@mails.tsinghua.edu.cn}},%\\[1mm] 
\\[10mm]
%\end{center}
%\\[10mm]
\centerline{$^{\star}$ \it
CAS Key Laboratory of Theoretical Physics,}
\centerline{\it
Institute of Theoretical Physics, Chinese Academy of Sciences,}
\centerline{\it Beijing 100190, P.\ R.\ China}
\vspace*{0.2cm}
\centerline{ $^\heartsuit $\it 
School of Physical Sciences, University of Chinese Academy
of Sciences,} 
\centerline{\it No.~19A Yuquan Road, Beijing 100049, P.\ R.\ China}
\vspace*{0.2cm}
\centerline{$^{\ast}$ \it
Department of Physics, Federal Urdu University of Arts, Science and Technology,}
\centerline{\it Karachi 75300, Pakistan}
\vspace*{0.2cm}
\centerline{$^{\diamond}$ \it Institute of Modern Physics, Tsinghua University, Beijing 100084, China}
\vspace*{1.cm}

\end{center}
%\vspace{1cm}

\begin{abstract}
\medskip

The null results of the LHC searches have put strong bounds on new physics scenario such as supersymmetry (SUSY). 
With the latest values of top quark mass and strong coupling, we study the upper bounds on the sfermion masses 
in Split-SUSY from the observed Higgs boson mass and electroweak (EW) vacuum stability. To be consistent with 
the observed Higgs mass, we find that the largest value of supersymmetry breaking scales $M_{S}$ for 
$\tan\beta=2$ and $\tan\beta=4$ are $\mathcal{O} (10^{3}\, {\rm TeV})$ and $\mathcal{O} (10^{1.5}\, {\rm TeV})$ respectively, 
thus putting an upper bound on the sfermion masses around $10^{3}\, {\rm TeV}$. In addition, 
 the Higgs quartic coupling becomes negative at much lower scale than the Standard Model (SM), 
and we extract the upper bound of $\mathcal{O}(10^{4}\, {\rm TeV})$ on the sfermion masses from EW vacuum stability. 
Therefore, we obtain the PeV-Scale Split-SUSY. The key point is the extra contributions to 
the Renormalization Group Equation (RGE) running from the couplings among Higgs boson, Higgsinos, and gauginos.
We briefly comment on the lifetime of gluinos in our study and compare it with current LHC observations. 
Additionally, we comment on the prospects of discovery of prompt gluinos in a 100 TeV proton-propton collider.

\end{abstract}

\bigskip
\bigskip

%% \begin{flushleft}
%% June 2004
%% \end{flushleft}

%%%%%%%%%%%%%%%%%%%%%%%%%%%%%%%%%%%%%%%%%%%%%%%%%%%%%%%%%%%%%%%
%\tableofcontents
%\vfill\eject

\section{Introduction}
\label{sec1}

The discovery of Higgs boson is the crowning achievement of particle physics \cite{Aad:2012tfa,CMS},
which completes the Standard Model (SM). Despite being one of the most successful scientific theory, the SM falls short in explaining some of the important issues. In particular, gauge coupling unification of the strong, weak and electromagnetic interactions of the fundamental particles is not possible. There is no explanation of the gauge hierarchy problem~\cite{ghp}. The SM also does not have a proper dark matter 
candidate.

The Supersymmetric Standard Model (SSM) is arguably the best bet to resolve all these issues in the SM.
Supersymmetry (SUSY) is an elegant scenario to solve the gauge hierarchy problem naturally.
The gauge coupling unification is a great triumph of the SSM~\cite{gaugeunification}.
The Minimal Supersymmetric Standard Model (MSSM) predicts that the Higgs boson mass ($m_{h}$) should be smaller than 135 GeV~\cite{mhiggs}. In addition, with $R$-parity conservation, the lightest supersymmetric particles (LSP) like neutralino etc is predicted 
to be an excellent dark matter particle~\cite{neutralinodarkmatter,darkmatterreviews}.

The discovery of the SM-like Higgs boson with mass $m_{h}\sim$ 125 GeV~\cite{Aad:2012tfa,CMS} requires the multi-TeV top squarks with small mixing or TeV-scale top squarks with large mixing~\cite{Carena:2011aa,Hahn:2013ria,Draper:2013oza,Bagnaschi:2014rsa,Vega:2015fna,Lee:2015uza,Bahl:2017aev}.  This observation questions the naturalness of the MSSM and specially the exsistance of the low scale ($\sim$ TeV) SUSY. Note that the TeV-scale SUSY is much more related to the ongoing LHC seraches and near future searches. On the other hand, LHC searches for SUSY or new physics have found no evidence of it~~\cite{Aaboud:2017vwy,Aaboud:2017hrg,Aaboud:2017aeu,Sirunyan:2018omt,Sirunyan:2018lul}. This situation has put the SUSY scenario under tension.

The SUSY electroweak fine-tuning problem  is a serious issue, and some promising and successful solutions can be found in the 
 literatures~\cite{Drees:2015aeo,Ding:2015epa,Baer:2015rja,Batell:2015fma,AbdusSalam:2015uba,Barducci:2015ffa,Cohen:2015ala,Fan:2014axa,Leggett:2014hha,Dimopoulos:2014aua,Gogoladze:2013wva,Kribs:2013lua,Gogoladze:2012yf}. Particularly, 
in an interesting scenario, known as Super-Natural SUSY~\cite{Leggett:2014hha, Du:2015una}, it was shown 
that no residual electroweak fine-tuning  (EWFT) is left in the MSSM if we employ the No-Scale supergravity boundary 
conditions~\cite{Cremmer:1983bf} and Giudice-Masiero (GM) mechanism~\cite{Giudice:1988yz} despite having   
relatively heavy spectra. One might think that the Super-Natural SUSY have a problem related to 
the Higgsino mass parameter $\mu$, which is generated by the GM mechanism and is proportional to 
the universal gaugino mass $M_{1/2}$. It should be noted that the ratio $M_{1/2}/\mu$ is of order one but cannot be determined 
as an exact number.  This problem, if it is,  can be addressed in the M-theory inspired the Next to MSSM (NMSSM) \cite{Li:2015dil}. 
Also, see~\cite{Baer:2017pba}, for more recent works related to naturalness within and beyond the MSSM. 

In this study we consider Split Supersymmetry (Split-SUSY) scenario proposed and developed 
in Refs.~\cite{Wells:2004di, ArkaniHamed:2004fb,Giudice:2004tc}. 
In this scenario, the fine-tuning of a light Higgs boson is accepted with ultra heavy sfermion masses, while the gauginos and Higgsinos 
are still around the TeV scale. Because spectrum has heavy scalars and relatively light fermions for sparticles, 
 it is called Split-SUSY. The fine-tuning of light Higgs is accepted on the same footing as of cosmological constant.
By having heavy sfermions, one can suppress additional CP-violation effects, excessive flavor-changing effects,
 and fast proton decays, etc. Also, one keeps gaugino and Higssino masses around the TeV scale, protected 
by the $R$ (chiral) symmetry. It was shown in \cite{Wells:2004di, ArkaniHamed:2004fb,Giudice:2004tc} that 
with super-heavy scalars and TeV scale fermions, gauge coupling unification is still preserved. 
Additionally, the lightest neutralino can still be
a dark matter candidate. In this way, one can still keep two very important virtues of SUSY scenario: 
 gauge coupling unification and  viable dark matter candidate. It is already known that the SUSY is a broken symmetry and that breakdown may happen at a high energy scale which is beyond the reach of the current LHC or future colliders such as FCC (Future Circular Collider)~\cite{Mangano:2019rww} or SppC (Super proton-proton Collider)~\cite{Tang:2015qga} searches. 
In this paper, we first employ the observed Higgs mass $m_{h}$ as a tool to predict the upper bound on the sfermion masses.
In numerical calculations, we vary top quark mass by $1\,\sigma$ from its central value, as well as vary $\tan\beta$ from 1 to 60.
With the latest values of top quark mass $M_{t}$ and strong coupling constant $\alpha_{3}$, and taking $m_{h}=$ 122, 125, and 127 GeV,
we obtain the SUSY breaking scale $M_{S}$ to be between $\sim 10^{3.5}$ to $10^{6}$ GeV, {\it i.e.}, 
$\mathcal{O}({\rm TeV})$ to $\mathcal{O}(10^3\,{\rm TeV})$. Thus, the upper bound on the sfermion mass is around $10^3$ TeV.
Moreover, by keeping $m_{h}=$ 123, 125, and 127 GeV, we calculate the SUSY breaking scale at which the Higgs quartic coupling $\lambda$ becomes negative, {\it i.e.}, from the electroweak (EW) vacuum stability bound. We show that $M_{S}$ turns out to be between 
$\mathcal{O}(10^3\,{\rm TeV})$ to $\mathcal{O}(10^4\,{\rm TeV})$, {\it i.e.}, the upper bound on sfermion mass is around $10^4$ TeV from the EW stability bound, which is much lower than
the corresponding scale around $10^{10}$ GeV in the SM~\cite{Degrassi:2012ry, Tang:2013bz, Buttazzo:2013uya}.  
Therefore, we obtain the PeV-Scale Split-SUSY from 
Higgs boson mass and EW vacuum stability. The main point is the extra contributions to 
the Renormalization Group Equation (RGE) running from  the couplings among Higgs boson, Higgsinos, and gauginos.
Since squarks are heavy, gluinos can be long lived depending on $M_S$. We briefly discuss 
the lifetime of gluinos in our calculations. It turns out that scenarios with $\tan\beta \gtrsim$ 5 have better chance of 
being discovered at the near future (33 TeV) or  
the long run (100 TeV) proton-proton colliders~\cite{Cohen:2013xda,Golling:2016gvc}.  

The remainder of the paper is organized as follows: In Section~\ref{sec2} we give calculations to find SUSY breaking scale. 
Section~\ref{sec3} contains discussion on the possible LHC SUSY searches for heavy scalars. 
We summarize our results and conclude in Section~\ref{sec4}.

\section{The Upper Bound on the Sfermion Masses from Higgs Boson Mass and Stability Bound}\label{sec2}

The discovery of Higgs boson at the LHC confirmed the SM as the low energy effective theory. But we have not found 
any new fundamental particle yet. Thus, we have even higher bounds on the new physics scenarios. In this situation, 
one can try to use Higgs boson mass $m_{h}$ and stability bound to put upper bounds on new physics (NP).  
In SM, the tree-level Higgs mass is defined as $m_{h}^{2}=2\lambda v^{2}$, where $\lambda$ is the SM Higgs self coupling (quartic coupling) which is a free parameter and $v$ is the vacuum expectation value. While in the SSMs,  $\lambda$ is not a free parameter, 
but it is related to other parameters of the model (we will discuss it more later). At tree level,
 the Higgs mass is bounded by $m_{Z}=91.187$ GeV, but to make it compatible with the LEP lower bound \cite{LEPHiggs} on the Higgs mass, 
we need to consider radiative correction~\cite{lista, Casas, Mariano, Haber}. For the Higgs boson with mass around 125~GeV, 
we need the multi-TeV top squarks with small mixing or TeV-scale top squarks with large mixing.
In  Split-SUSY, we can obtain  $\lambda\left(M_{S}\right)$, which is defined at the SUSY breaking scale $M_S$ 
where sfermions are decoupled. We then calculate the low energy Higgs quartic coupling via 
the RGE running, and get the low energy effective field theory \cite{Casas}. 
Because enhancing $\lambda$ will increase the Higgs boson mass,  the upper and lower bounds on $m_{h}$ can put 
the corresponding bounds on $M_S$ as well.
The tree level Higgs quartic coupling is defined at the SUSY breaking scale $M_{S}$ \cite{ArkaniHamed:2004fb,Giudice:2004tc,Barger} as
%%%%%%%%%%%%%%%%%%%%
\begin{equation}
\lambda(M_{S})=\frac{g_{1}^{2}(M_S)+g_{2}^{2}(M_S)}{4}\cos 2\beta,
\end{equation}
%%%%%%%%%%%%%%%%%%%%%
where $g_{2}$ and $g_{1}=\sqrt{\frac{5}{3}}g_{Y}$ are $SU(2)$ and $U(1)_{Y}$ gauge couplings, repectively. The parameter $\beta$ is given as $\tan\beta\equiv \frac{v_{u}}{v_{d}}$ with $v_{u}$ and $v_{d}$ are vaccum expectations values of the Higgs doublets $H_{u}$ and $H_d$, rexpectively. Above $M_{S}$, SUSY is restored and below $M_{S}$ the SUSY is broken. In order to predict the Higgs mass, below $M_S$, we use the split-SUSY  2-loop renormalization evolution (RGE) along with 2-loop gauginos and higgsinos contributions and one-loop RGE running for the SM fermion Yukawa couplings (for RGEs, see appendix of \cite{Giudice:2004tc}).

In our numerical calculations we use the fine structure constant $\alpha_{EM}$ , weak mixing angle $\theta_{W}$ at $M_{Z}$ as follows \cite{Tanabashi:2018oca}:
%%%%%%%%%%%%%%%%%%%%%%%%%%%%%%%%%%%%%%%%%
\begin{equation}
\sin^{2}{\theta_{W}}=0.22332\pm 0.00007 \qquad \alpha^{-1}_{EM}=137.03599.
\end{equation}
%%%%%%%%%%%%%%%%%%%%%%%%%%%%%%%%%%%%%%%
The top quark pole mass and the strong coupling constant are very important parameters in our calculations and we take their values respectively as: 
\begin{equation}
M_ {t} = 173.34\pm 0.76 \,{\rm GeV} \text{\cite{ATLAS:2014wva}},\qquad \alpha_{3}(M_{Z} )= 0.1187 \pm 0.0016  \text{\cite{Tanabashi:2018oca}}.
\end{equation}
We also use the one-loop effective Higgs potential
with top quark radiative corrections and calculate the Higgs boson mass by minimizing the
effective potential given in \cite{Huo:2010fi}

%%%%%%%%%%%%%%%%%%%%%%%%%%%%%%%%%%%%
%%%%%%%%%%%%%%%%%%%%%%%%%%%%%%%%%
\begin{figure}
		\centering	
		\includegraphics[width=.5\linewidth]{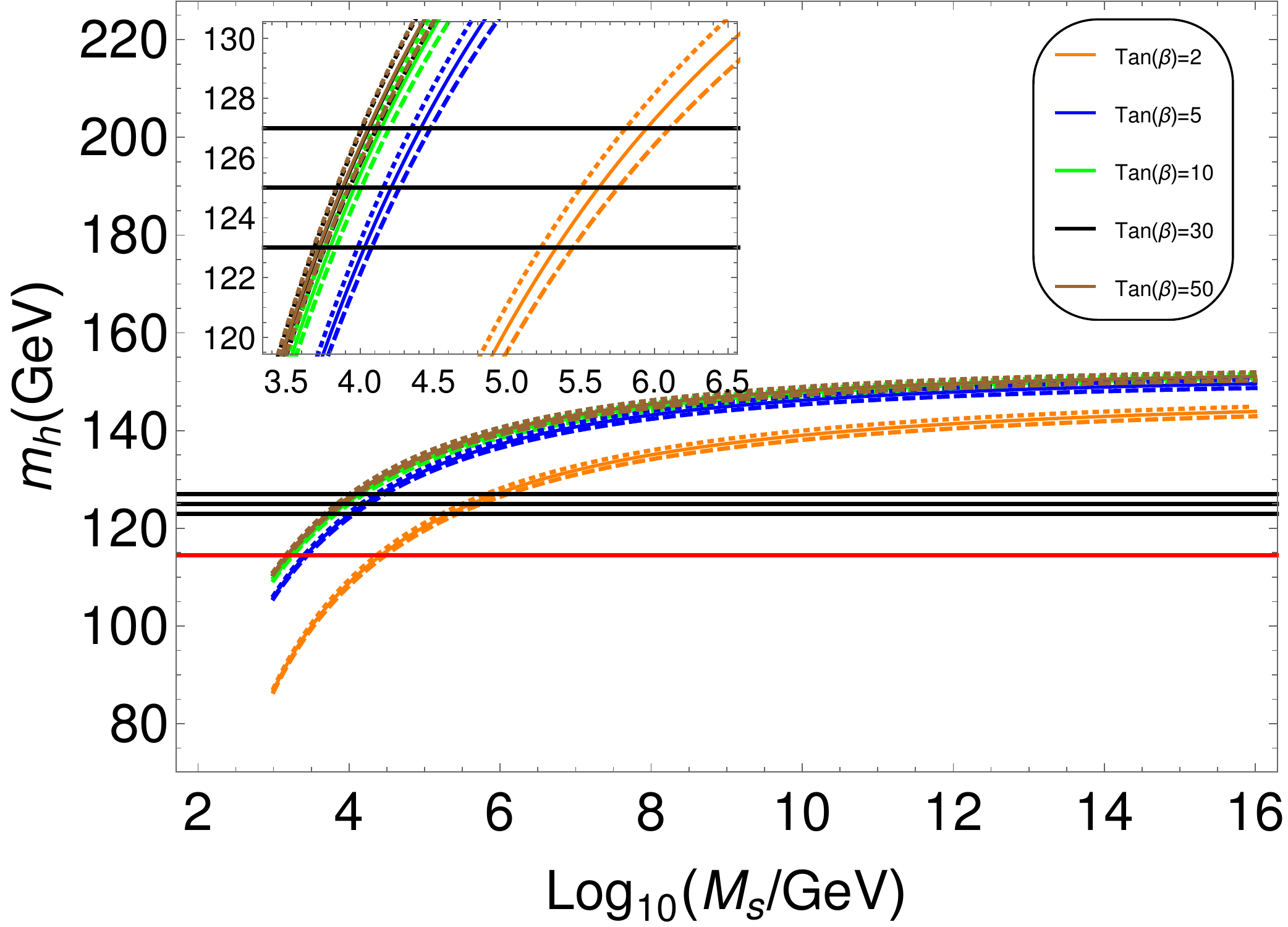}
%		\caption{\small{$m_{s}$ vs $m_{h}(GeV)$ }}.
		\label{3d2}
	
	\caption{\small{The predicted Higgs boson mass $m_{h}$ versus $M_{S}$ in Split-SUSY for a constant $\tan \beta$ along the curves. The horizontal red line is the LEP lower bound 114.5 GeV. The solid, dashed and dotted curves represent calculations with $M_{t}$ , $ M_ {t} - \delta M_{t}$ and $ M_ {t} + \delta M_{t}$, respectively.}}
	\label{mhMs}
\end{figure}
%%%%%%%%%%%%%%%%%%%%%%%%%%%%%%%%%%%%%%%%%%%%
%%%%%%%%%%%%%%%%%%%%%%%%%%%%%%%%%%%%%%%%%%%%
\begin{equation}
V_{eff}=m_{h}^{2}H H^{\dagger} +\frac{\lambda}{2}\left(H H^{\dagger}\right)^{2}-\frac{3}{16}h_{t}^{4}\left(H H^{\dagger}\right)^{2}\left[Log\left(\frac{h(t)^{2}H H^{\dagger}}{Q^{2}}\right)-\frac{3}{2}\right],
\end{equation}
%%%%%%%%%%%%%%%%%%%%%%%%%%%%%%%%%%%%%%%%%%
where  $m^2_{h}$ is the squared Higgs boson mass, $h_{t}$ is the top quark Yukawa coupling from
$m_{t} = h_{t} v_{u}$, and the scale $Q$ is chosen to be at the Higgs boson mass. For the $\overline{MS}$ top quark mass $m_t$, we use the two-loop corrected value, which is related to the top quark pole mass
$M_t$ \cite{Huo:2010fi},

%%%%%%%%%%%%%%%%%%%%%%%%%%%%%%%%%%%%%%%%%%%%%%%%%%%
%%%%%%%%%%%%%%%%%%%%%%%%%%%%%%%%%%%%%%%%%%%%%%%%%%%%%%

\begin{eqnarray}
M_t&=&m_t(m_t)\left\{1+\frac{4\alpha_3(m_t)}{3\pi}+\left[13.4434-1.0414\sum_{k=1}^5(1-\frac{4}{3}\frac{m_k}{m_t})\right]\left[\frac{\alpha_3(m_t)}{\pi}\right]^2 \right\},
\end{eqnarray}
%%%%%%%%%%%%%%%%%%%%%%%%%%%%%%%%%%%%%%%
where $m_{k}$ represents other quark masses. Additionally for $\alpha_{3}(m_{t})$ we use two-loop RGE running.
We display our calculations in Figure~\ref{mhMs} which shows the relation between the light mass $m_{h}$ and $M_{S}$ for different value of $\rm tan\beta$. The horizontal red line represents the LEP bound on the Higgs mass, while horizontal black lines represent $m_{h}=$ 123 GeV, 125 GeV and 127 GeV. We allow variation of $\pm$ 2 GeV from $m_h$= 125 GeV due to uncertainty in theoretical calculations \cite{Allanach:2004rh}. Solid curves represent calculations with $M_{t}$ while dotted (dashed) curves depict calculations with $M_{t} \pm\delta M_{t}$ with $\delta M_{t}=$ 0.76 GeV.
As we have stated above, the quartic coupling depends on $\cos^{2}2\beta$, so the Higgs mass remains constant for large value of $\rm tan\beta$. This trend can clearly be seen in Figure~\ref{mhMs} that is for larger values of $\rm \tan\beta$ ($\geq$ 5), all the curves come near to each other. This shows that one can not produce any arbitrary value of the Higgs mass by picking any $M_{S}$ value for a given $\rm tan\beta$. Therefore, for fixed $\rm tan\beta$, we have an upper bound on the Higgs mass which becomes $\sim 147$ GeV.

As discussed above, in split-SUSY, all the sfermions have masses of the same order as of SUSY breaking scale $M_S$,
  but gauginos and Higgsions can be light. From the Figure~\ref{mhMs} we see that for $m_{h}=$ 123 GeV and $m_{h}=$ 127 GeV the corresponding values for $M_{S}$ is $10^{5.2}$ GeV to $10^{6.2}$ GeV respectively for $\rm \tan\beta =$ 2. For $\rm \tan\beta =$ 50, we have $M_{S}$ ranges from $10^{3.6}$ GeV and $10^{4.2}$ GeV. Thus, we show that the lower bound on sfermion masses is $\sim$ 1 TeV and upper bound is 
about $10^3$ TeV. We note that since $M_{S}\lesssim 10^{4.5}$ GeV for $\tan\beta \gtrsim$ 4, we have more chance to probe Split-SUSY scenario 
at the future collider \cite{Cohen:2013xda,Golling:2016gvc}. We will discuss it more later in the next section. It is very important to note that how the observed Higgs boson mass restricts the possible values of the SUSY breaking scale $M_{S}$ and which is really astonishing. We also like to make a comment here. Our lower and upper values of $M_{S}$ is somewhat different with \cite{ArkaniHamed:2004fb,Giudice:2004tc}. We note that this is because of the $M_{t}$ and $\alpha_{3}$ we use in our calculations. 

%%%%%%%%%%%%%%%%%%%%%%%%%%%%%%%%%
%%%%%%%%%%%%%%%%%%%%%%%%%%%%%%%%%%%%%%%%%%%%%5
\begin{figure}[t!]
\centering

\includegraphics[width=0.5\textwidth]{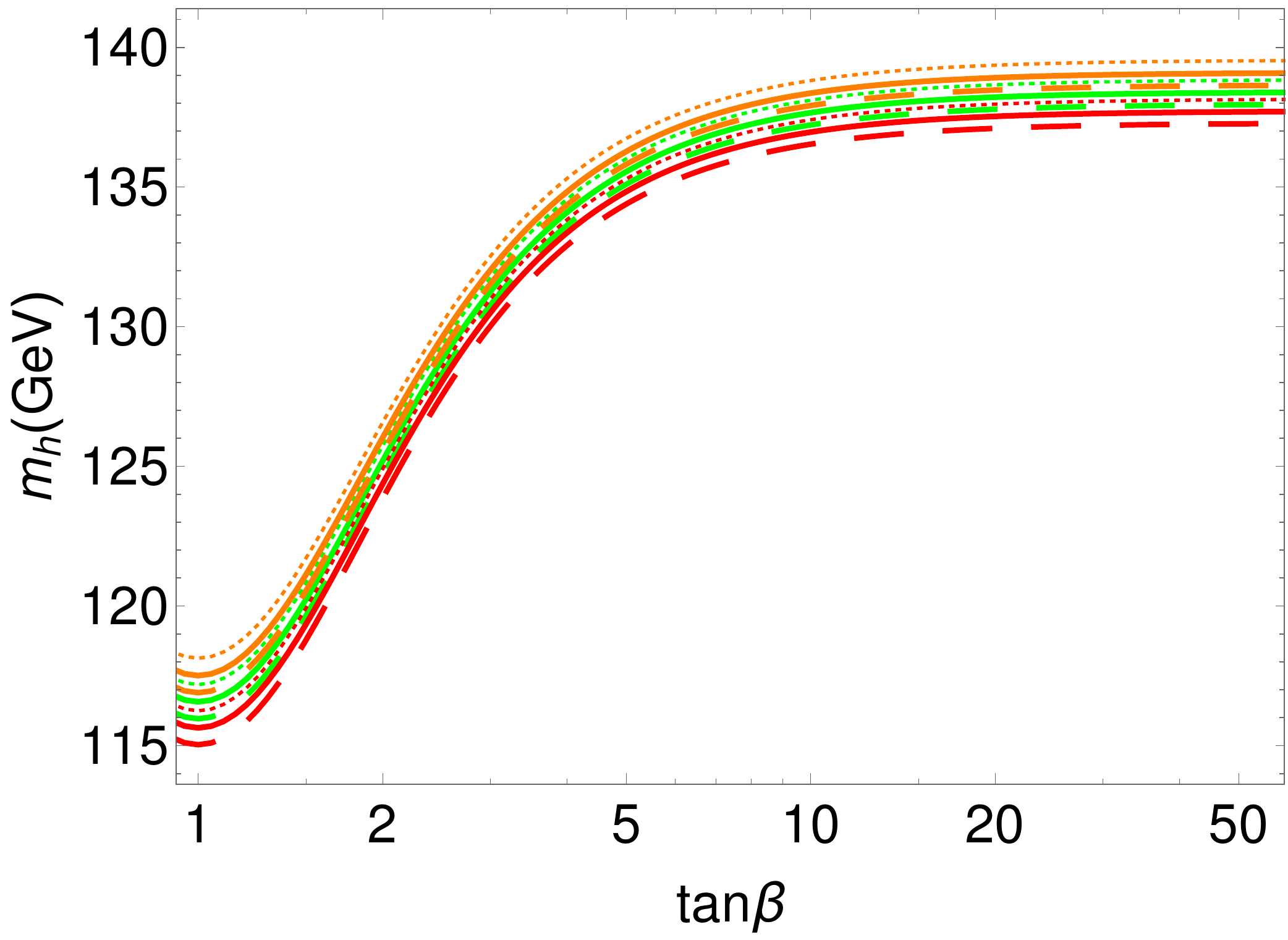}

\caption{\small{ The predicted Higgs boson mass $m_{h}$ versus $\tan \beta$ in the Split-SUSY with $M_{S} \left(10^{5.2}\right)$ GeV. The orange, green, and red curves are for $M_{t} + \delta M_{t}$, $M_t$, and $M_{t} - \delta M_{t}$, respectively. 
The dashed (dotted) curves are for $\alpha_{3} \pm \delta \alpha_{3}$, and the solid ones are for $\alpha_{3}$.}}
\label{mhtanb}
\end{figure}
%\\

%%%%%%%%%%%%%%%%%%%%%%%%%%%%%%
Figure~\ref{mhtanb} shows the relation between the Higgs mass $m_{h}$ and $\tan\beta$ for different values of $\alpha_{3}$ and top quark mass $M_{t}$ with $M_{S}=\left(10^{5.2}\right)$ GeV. The orange, green and red curves are for $M_{t} + \delta M_{t}$, $M_t$ and $M_{t} - \delta M_{t}$ respectively. The dashed (dotted) curves are for $\alpha_{3} \pm \delta \alpha_{3}$ and the solid ones are for $\alpha_{3}$.
The solid curves clearly show that when we increase the top quark mass the Higgs mass also increases and vice versa. In addition to it, one can  see that when we decrease the $\alpha_{3}$ by $1\sigma$ ($\delta \alpha_{3}$), the Higgs mass increases which is represented by dotted curves, similarly when we increase the $\alpha_{3}$ by $1\sigma$, the Higgs mass decreases which is shown by dashed curves. In our calculations the predicted Higgs boson mass ranges from $114$ to $139$ GeV for the variation in $\tan\beta$ from 1 to 60.

%%%%%%%%%%%%%%%%%%%%%%%%%%%%%%%%%
\begin{figure}\centering
%\begin{figure}[htp]
%\centering
\subfiguretopcaptrue

\subfigure{
\includegraphics[totalheight=5.5cm,width=7.cm]{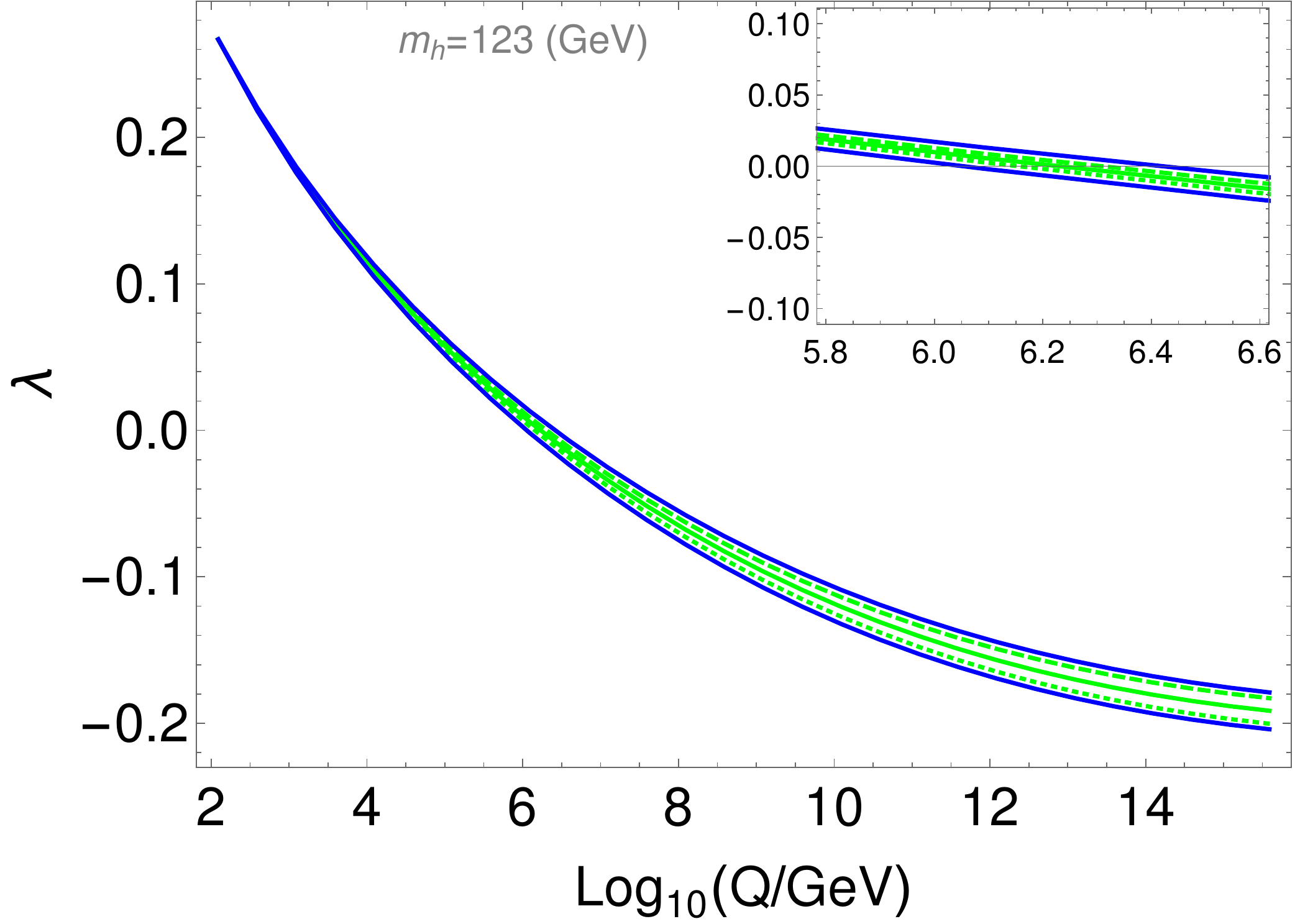}
}
\subfigure{
\includegraphics[totalheight=5.5cm,width=7.cm]{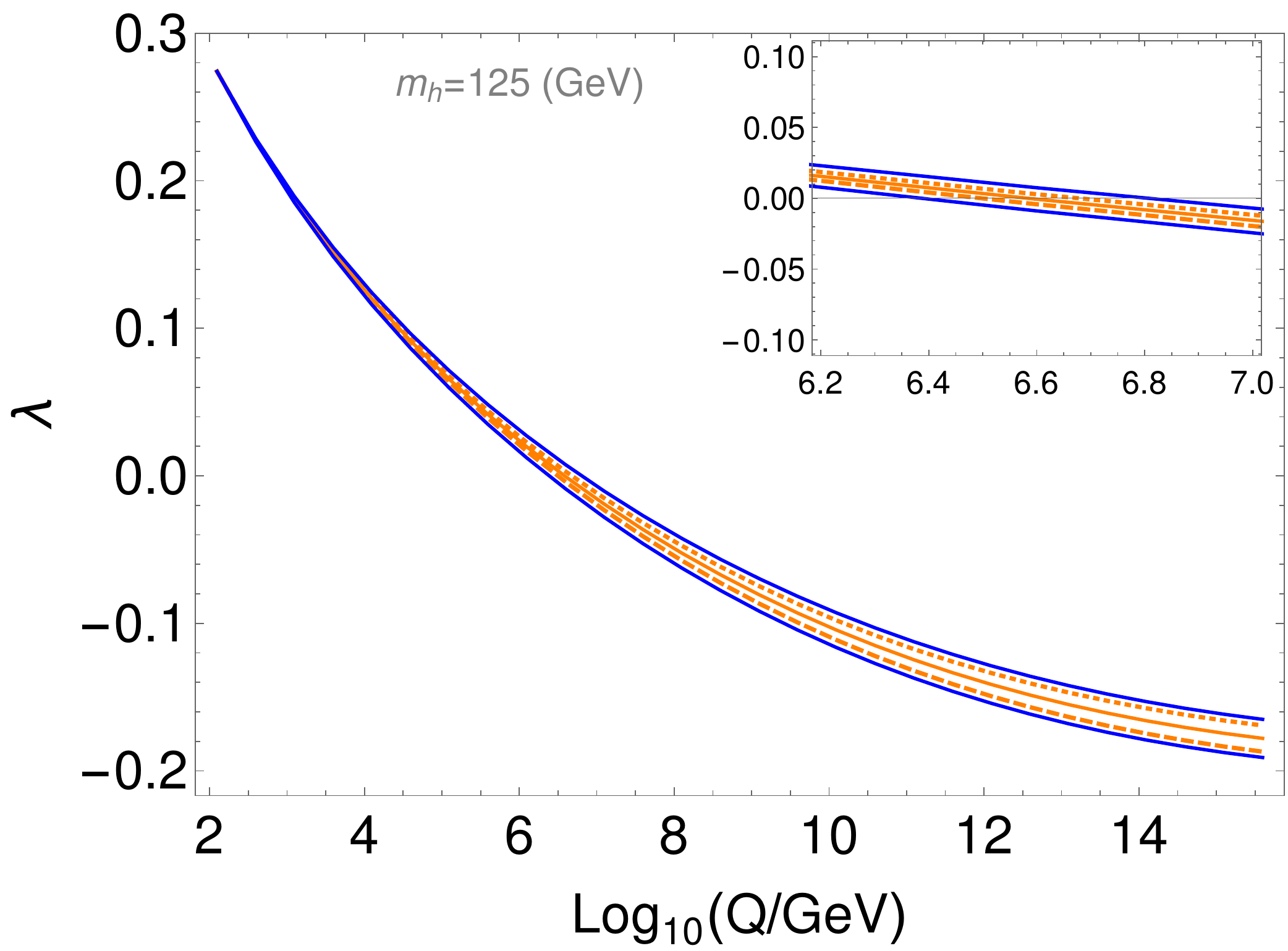}
}
\subfigure{
\includegraphics[totalheight=5.5cm,width=7.cm]{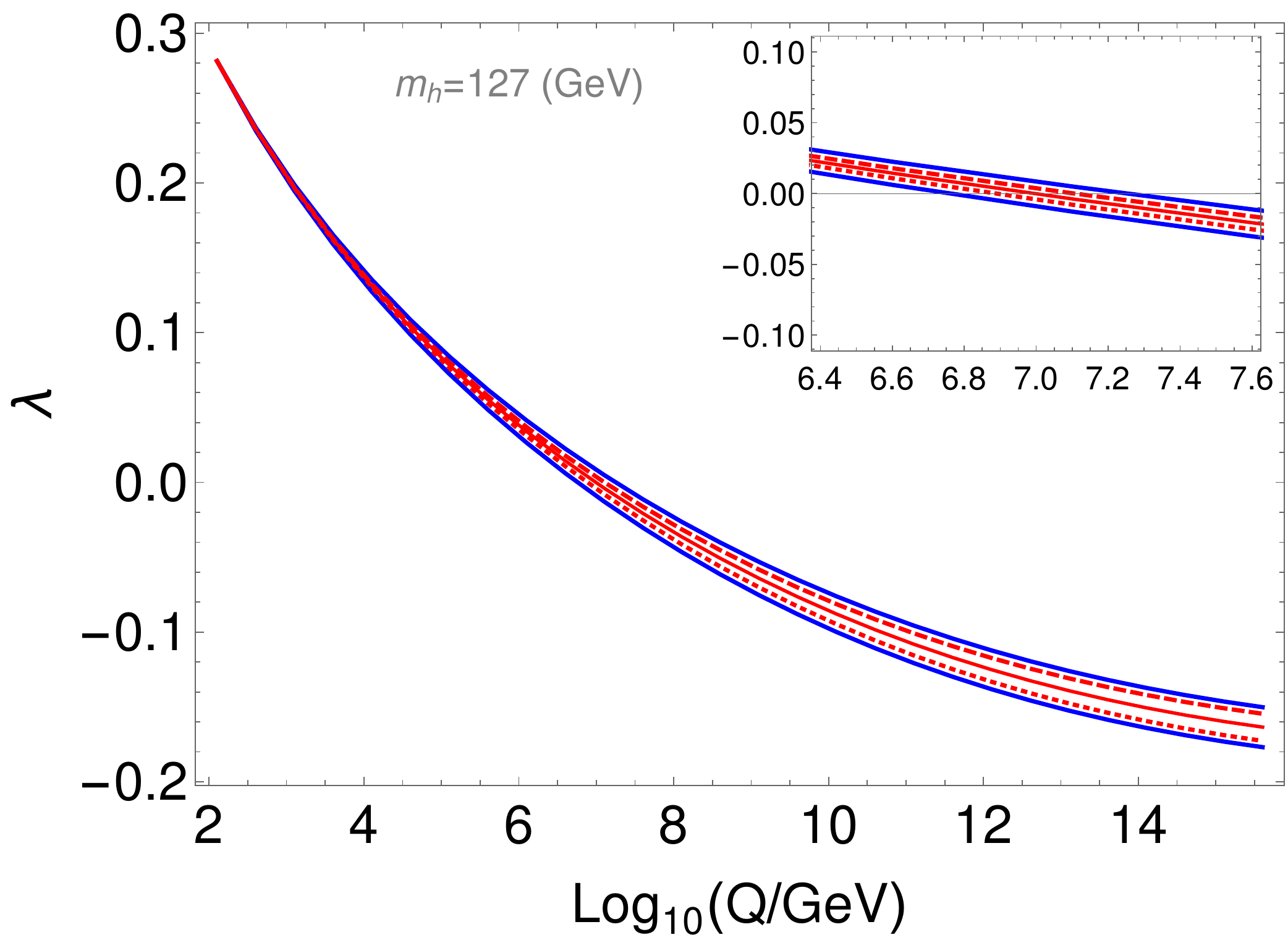}
}

\caption{\small{Figure shows the running of quartic coupling $\lambda(t)$ versus $Q$ for Split-SUSY scenario.  The central solid curves in each plot correspond to $M_{t}=173.34$ GeV and $\alpha_{3}=0.1187$. The dashed curves are for $\alpha_{3}+\delta \alpha_{3}$,  and the dotted ones are for  $\alpha_{3} - \delta \alpha_{3}$. The outer blue upper and lower curves correspond to $ M_ {t} + \delta M_{t}$ and $ M_ {t} - \delta M_{t}$, respectively.}}
		\label{lamQ}
\end{figure}
%%%%%%%%%%%%%%%%%%%%%%%%%%%%%%%%%%%%%%%%%%%%%%%%%%%%%%%%%%%%%%%%%%%%%%%%

It is well-known that if the Higgs mass is below $127$ GeV, the radiative corrections from the top quark make the Higgs quartic coupling $\lambda$ negative below the Planck scale $(M_{\mathrm{Pl}})$ and as a result the Higgs potential becomes unstable due to quantum tunneling.
This is called stability problem. In the Split-SUSY, $M_S$ must be below the scale at which the Higgs quartic coupling becomes negative,
so the upper bounds on $M_S$ can be obtained. We consider two-loop quartic couplings RGEs \cite{Chen:2018ucf}. Figure~\ref{lamQ} shows the running of quartic couplings $\lambda$ along energy scale $Q$. We show our calculations of $\lambda$ for $m_{h}=$ 123 GeV, 125 GeV and 127 GeV. The central solid curves in each plot correspond to the central values of $M_t$ and $\alpha_3$. The dashed (dotted) curves represent $\alpha_{3} \pm \delta \alpha_{3}$ and outer upper (lower) curves depict variation of $ M_ {t} \pm \delta M_{t}$. From the curves we see that
the quartic couplings will increase or decrease when we increase or decrease the top quark mass. But we note that when we increase $\alpha_{3}$ the quartic coupling decreases and vice versa which is opposite to the case of variation in top quark mass. Plots show that quartic coupling cut the x-axis at $Q\simeq10^{6.1}$ GeV for $m_{h}=$ 123 GeV and $Q\simeq 10^{7.3}$ GeV for $m_{h}=$ 127 GeV which give the lower and upper bounds on the energy scale at which vacuum become unstable. 
The vacuum stability bounds are related to the SUSY breaking scale too. Thus, we have the lower and upper bounds 
on scalar sparticle masses of about 100 TeV to 1000 TeV, and then we can call it ``PeV-Scale Split Supersymmetry''.

\section{LHC Searches}\label{sec3}

In a hadron collider the standard signatures of SUSY consist of multi-jet and multi-lepton final states with missing transverse energy $\slash{E_T}$ \cite{baer,Martin:1997ns}. The underlying physics typically involves pair production of new heavy coloured particles (squarks and gluinos), which cascade-decay into the LSP. In Split-SUSY, since all the 
squarks are heavy, these cascade decays will essentially not occur. The only available states that may be light are bino($M_{1}$), wino ($M_{2}$), Higgsinos and gluinos ($M_3$). The requirement of viable dark matter particle 
may force bino, wino and Higgsinos to be around TeV \cite{Wang:2005kf,Masiero:2004ft,Pierce:2004mk}.  Also see ref.~\cite{Zhu:2004ei} for the decays of charginos in Split-SUSY.

Gluinos are supposed to be the smoking gun signature for SUSY. The absence of TeV-scale squarks in Split-SUSY effect the production and decays of gluinos as compared to the MSSM. For example, in split-SUSY we can still pair produce gluinos via $gg$ and $q{\bar q}$ annihilations and unlike the MSSM, squark exchange production channel is negligible \cite{Hewett:2004nw}. So the production rate of gluino is some what smaller in split-SUSY as compared to the MSSM.

In Split-SUSY the available decay channels are $\tilde g \rightarrow q{\bar q}{\tilde \chi^{0}}$, $\tilde g \rightarrow q q^{\prime}{\tilde \chi^{\pm}}$ and loop induced channel $\tilde g \rightarrow \gamma{\tilde \chi^{0}}$. Since these decays involved squarks, gluino can be long lived \cite{Hewett:2004nw,Kilian:2004uj,Gambino:2005eh}. 
The latest LHC direct searches have already excluded gluinos lighter than 2 TeV \cite{ATLAS:2017cjl,Sirunyan:2017cwe,Sirunyan:2017kqq}. In addition to it, the LHC searches for long lived massive particles have also put limits on long lived gluinos. In Ref.~\cite{Aaboud:2017iio} the ATLAS Collaboration provides the exclusion limits on the production of long-lived gluinos with masses up to 2.37 TeV and lifetimes of $\mathcal{O}(10^{-2})-\mathcal{O}(10)$ nano seconds in a simplified model inspired by Split-SUSY. In another search \cite{Sirunyan:2018vjp} the CMS Collaboration provides sensitivity to the simplified models
inspired by Split-SUSY that involve the production and decay of long-lived gluinos.
They consider values of the proper decay length $c\tau_{0}$ from $10^{-3}$ to $10^5$ mm. Gluino masses up to 1750 and 900 GeV are probed for
$c\tau_{0}$ = 1 mm and for the metastable state, respectively. The sensitivity is moderately
dependent on model assumptions for $c \tau_{0}\gtrsim$ 1 m. The gluino lifetime in terms of $M_{S}$ and gluino mass $m_{\tilde g}$ is given as \cite{Gambino:2005eh}
\begin{equation}
\tau_{\tilde g}\approx \rm{4 \, sec} \times \left(\frac{M_{S}}{10^{9} \,\rm{GeV}}\right)^{4} \times \left( \frac{\rm{1 TeV}}{m_{\tilde g}}\right)^{5}.
\label{glulifetime}
\end{equation}
   
\noindent A part from usual gluino (prompt) signal, long lived gluinos can also produce interesting signals for collider searches. The gluinos with lifetime around $\sim 10^{-12}$ seconds will most likely to decay in the detector and can be probed by $\rm multi-jet+ \EmissT$  but with displaced vertices \cite{Hewett:2004nw}.
On the other hand,  gluinos with lifetimes between $10^{-12}$ to $10^{-7}$ seconds will decay in the bulk of the detector. Finding such gluinos can be a problem as QCD background may hide them.
The gluinos with $\tau_{\tilde g}\gtrsim 10^{-7}$ seconds are expected to decay outside of the detector. In this case they will appear to be effectively stable, and search strategies for heavy
stable particles need to be employed \cite{Hewett:2004nw,Perl:2001xi}. In order to get a rough estimate of the gluino lifetime in our case, we assume $m_{\tilde g}$= 2.5 TeV.  For $M_{S}\sim 10^{3}$ GeV, $\tau_{\tilde g}\sim 10^{-26}$ seconds which corresponds to prompt gluino. Such a scenario can be realized when gluino decays into light flavour and neutralino ($\tilde g \rightarrow q {\bar{q}}{\tilde \chi_{1}^{0}}$) and gluino decays into heavy flavor and neutralino ($\tilde g \rightarrow t {\bar{t}}{\tilde \chi_{1}^{0}}$). It was shown in Refs.~\cite{Cohen:2013xda,Hewett:2004nw} that assuming $m_{\tilde \chi_{1}^{0}}<$ 1 TeV, for the first case ($\tilde g \rightarrow q {\bar{q}}{\tilde \chi_{1}^{0}}$), gluinos lighter than 11 TeV can be discovered,
 and for the second case ($\tilde g \rightarrow q {\bar{q}}{\tilde \chi_{1}^{0}}$), gluinos lighter 8 TeV can be discovered at
a 100 TeV proton-proton collider. On the other hand, for $M_{S}$ values of $10^{6}$ and $10^{7}$ GeV, 
the lifetime of gluino is $\sim 10^{-14}$ and $10^{-10}$ seconds, respectively.  
We hope that the future analysis at higher energies become more  effective,
 and we will see some hints for such gluinos and eventually the SUSY breaking scale $M_{S}$.

\section{Conclusions}
\label{sec4}

Taking the latest values of top quark mass and strong coupling, we have studied the upper bounds on the sfermion masses 
in Split-SUSY from the observed Higgs boson mass and EW vacuum stability. 
Varying top quark mass by $1\,\sigma$ from its central value and calculating the Higgs boson mass for various values of $\tan\beta$, we first found that for $m_{h}=$ 122, 125, and 127 GeV, the SUSY breaking scale turns out to be between  
$\mathcal{O}({\rm TeV})$ to $\mathcal{O}(1000\,{\rm TeV})$, 
thus putting an upper bound on the sfermion masses around $10^{3}\, \rm TeV$. 
In addition, for $m_{h}=$ 123, 125, and 127 GeV, we showed that the 
SUSY breaking scale at which the Higgs quartic coupling $\lambda$ becomes negative is 
 between $\mathcal{O}(10^{3}\,{\rm TeV})$ to $\mathcal{O}(10^{4}\,{\rm TeV})$. 
So we extract the upper bound of $\mathcal{O}(10^{4}\, {\rm TeV})$ on the sfermion masses from EW vacuum stability. 
Therefore, we obtain the PeV-Scale Split-SUSY. The key point is the extra contributions to the RGE running
 from  the couplings among Higgs boson, Higgsinos, and gauginos.
 Since squarks are heavy, gluinos can be long lived depending on $M_S$, and we discussed
the lifetime of gluinos. It turns out that the scenarios with $\tan\beta \gtrsim$ 5 have better chance of 
being discovered at the near future (33 TeV) or the long run (100 TeV) proton-proton colliders.

\section{Acknowledgments}

This research was supported by the Projects 11475238, 11647601, and 11875062 supported by the 
National Natural Science Foundation of China, and by the Key Research Program of Frontier Science, CAS.WA  was  supported  by  the  CAS-TWAS  Presidents  Fellowship  Programme, China.

%%%%%%%%%%%%%%%
%%%%%%%%%%%%%%%%%%%%%%%%%%%%%%%%%%%%%%%%%%%%%%%%%%%%%%%%%%%%%%%%%%%%%%%%

%%%%%%%%%%%%%%%%%%%
\end{document}